\documentstyle[amssymb,aps,multicol,prl]{revtex}
\draft

\begin{document}
\title{ Dynamic creation and annihilation of metastable vortex
phase as a source of excess noise}
\author{Y. Paltiel$^1$, G. Jung$^{1,2,*}$, Y. Myasoedov$^1$, M. L. Rappaport$^1$,
E. Zeldov$^1$, M. J. Higgins$^3$, and S. Bhattacharya$^{3,4}$}
\address{$^1$Department of Condensed Matter Physics, Weizmann Institute of
Science, Rehovot 76100, Israel}
\address{$^2$Department of Physics, Ben-Gurion University of the Negev, Beer-Sheva 84105, Israel}
\address{$^3$NEC Research Institute, 4 Independence Way, Princeton, New Jersey 08540}
\address{$^4$Tata Institute of Fundamental Research, Mumbai-400005, India}

\date{\today}
\maketitle

\begin{abstract}
The large increase in voltage noise, commonly observed in the
vicinity of the peak-effect in superconductors, is ascribed to a
novel noise mechanism. A strongly pinned metastable disordered
vortex phase, which is randomly generated at the edges and
annealed into ordered phase in the bulk, causes large fluctuations
in the integrated critical current of the sample. The excess noise
due to this dynamic admixture of two distinct phases is found to
display pronounced reentrant behavior. In the Corbino geometry the
injection of the metastable phase is prevented and, accordingly,
the excess noise disappears.

\end{abstract}


\begin{multicols}{2}

The appearance of large noise with the onset of motion of a
condensate in the presence of random pinning potential, has been
studied extensively in incommensurate charge density waves
\cite{cdw}, Wigner crystals in two-dimensional electron gas
\cite{li}, and most notably, in vortex matter in type-II
superconductors
\cite{clemrev,yeh,placais,woltgens,marley95,merthew,rabin,danna,safar,okuma,vinokur}.
In all cases, the noise is thought to reflect spatio-temporal
irregularities of the moving condensate due to its interaction
with the underlying pinning potential, though its precise origin
remains obscure and controversial. The voltage noise due to vortex
motion in a current-biased superconductor is generally referred to
as flux-flow noise for which various mechanisms have been
considered (for an early review see \cite{clemrev}). They include
vortex shot noise and the associated density fluctuations
\cite{clemrev,yeh}, velocity fluctuations resulting from
vortex-pin interactions \cite{clemrev} or turbulent flow of
surface currents \cite{placais}, critical slowing down of vortex
dynamics \cite{woltgens}, and several suggestions
\cite{marley95,merthew,rabin,danna,safar,okuma} and numerical
simulations \cite{vinokur} of various plastic vortex flow
mechanisms. Each of these mechanisms may make a substantial
contribution to the total measured noise. Yet the puzzling
observation, which has no satisfactory explanation, is that in a
specific and narrow region of the $H-T$ phase diagram the noise is
enhanced drastically. This {\em excess noise} exceeds the usual
flux-flow noise level by orders of magnitude
\cite{clemrev,marley95,merthew,rabin,danna,safar}. In low-$T_c$
superconductors the excess noise occurs in the vicinity of the
peak effect (PE) below $H_{c2}$, where the critical current $I_c$
anomalously increases with field
\cite{clemrev,marley95,merthew,rabin}. In high-$T_c$
superconductors (HTS) similar noise enhancement was found in the
vicinity of the melting or order-disorder transitions
\cite{danna,safar,kwok,gordeev}. This low frequency excess noise
is apparently inconsistent with the common flux-flow noise
mechanisms due to its unusually high amplitude, strong field and
current dependence, and non-Gaussian character
\cite{marley95,merthew,rabin,danna}.

In this paper we demonstrate that the excess noise does not result
from any previously known flux-flow noise mechanism, but rather is
due to a conceptually different phenomenon of random creation and
annihilation of a metastable phase. The conventional models
consider various types of irregularities and defects in the vortex
lattice which are perturbations within an otherwise single
thermodynamic phase of the vortex matter. In contrast, it has
recently been suggested that the PE in low-$T_c$ superconductors,
as well as the second magnetization peak in HTS, reflects a
first-order phase transition between two distinct thermodynamic
phases of the vortex matter \cite{yosprl,maxim,nurit,kierfeld}: an
ordered phase (OP, or Bragg glass \cite{gl}), and an amorphous
disordered phase (DP). The DP is the equilibrium thermodynamic
phase above the peak field of the PE, whereas the OP is the
equilibrium phase below the peak field. However, the DP can be
also present as a `wrong' {\em metastable} supercooled phase below
the PE instead of the equilibrium OP \cite{maxim,yosna,hend96}. As
shown below the excess noise results from the {\em dynamic
coexistence of these two distinct phases of the vortex matter}. In
the lower part of the PE the random penetration of vortices
through the sample edges does not just create some defects in the
OP, but rather generates a distinct, albeit metastable, phase. The
random generation and annihilation of this strongly pinned
metastable DP is shown to be the cause of large fluctuations in
the instantaneous $I_c$ of the sample, leading to greatly enhanced
voltage noise. We demonstrate accordingly that the excess noise
can be {\em eliminated} by preventing the formation of the `wrong'
phase in the Corbino disk geometry. In addition, the excess noise,
which was previously observed only in the vicinity of the PE where
$I_c$ increases with field $H$, is found also at low fields, where
$I_c$ {\em decreases} with $H$.

To investigate the origin of the excess noise we have studied
2H-NbSe$_2$ crystals using a special contact configuration (Fig.
1b, inset) that enables measurements in both the Corbino and
strip-like geometry in the same crystal \cite{yosprl}. By applying
the current to the +S,-S contacts, the vortices penetrate through
the edge and flow across the sample, similarly to the standard
configuration. In contrast, by applying the current to the +C,-C
contacts, the vortices circulate in the bulk without crossing the
edges, as in a Corbino disk. In both configurations the voltage
and the corresponding noise are measured across the same contacts
+V,-V. The distance between the voltage contacts is 0.15 mm and
the diameter of the Corbino is 1.1 mm. The data presented here
were obtained on a Fe-doped (200 ppm) NbSe$_2$ single crystal 2.2
$\times $ 1.5 $\times $ 0.04 mm$^{3}$ with $T_c$=5.7 K
\cite{yosprl}. Similar results were obtained on a number of
additional crystals.

The inset of Fig. 1a shows the voltage response $V$ vs. the field
$H \|$c-axis at 4.4 K in the Corbino and strip configurations. The
applied current $I=$ 26 mA in the strip configuration and 34 mA in
the Corbino. Since the current density varies across the sample,
this difference in $I$ is chosen to give the same average current
density between +V,-V contacts in the two geometries. As a result,
the measured $V$ and the corresponding vortex velocity are
identical at high fields. Upon decreasing the field from above
$H_{c2}(T)$, the voltage decreases rapidly and vanishes in the PE
region (4 to 8 kOe, Fig. 1a inset) where $I_c$ of the sample is
large due to the presence of the strongly pinned DP. The voltage
becomes finite at intermediate fields before vanishing again at $H
\lesssim$ 1.2 kOe. We shall concentrate on the intermediate fields
where the excess noise appears and where significantly different
response in the two geometries is observed.

In the Corbino, $V$ increases linearly with $H$ (Fig. 1a) in
accord with the well-known flux-flow behavior in undoped NbSe$_2$
\cite{bhatta}, indicating that the lattice is in the OP. The
voltage vanishes abruptly at both high and low fields. The
high-field point $H_{DT}^H$ is the sharp disorder-driven
transition at the PE between the equilibrium OP and the
equilibrium DP, whereas the low-field point $H_{DT}^L$ is the
reentrant disorder-driven transition \cite{yosprl,ghosh}. The
strip configuration displays qualitatively different behavior
(Fig. 1a). In contrast to Corbino, no sharp transitions are
observed and the voltage remains vanishingly small below
$H_{DT}^H$ as well as above $H_{DT}^L$. At intermediate fields $V$
increases, but it is still significantly suppressed relative to
the Corbino response.

Figure 1b shows the low frequency noise in the two geometries. The
large noise in the strip was previously observed only in the lower
part of the PE \cite{marley95,merthew,rabin}. Here we find, for
the first time, two peaks in the noise. The noise of the strip is
maximal where voltage response is strongly suppressed, while the
noise is reduced in the central region, where $V$ of the strip
recovers. The most important observation in Fig. 1b, however, is
that the large excess noise that is found in the usual strip
configurations is absent in the Corbino geometry, and only two
small peaks remain at $H_{DT}^H$ and $H_{DT}^L$. This is a first
demonstration that the excess noise depends dramatically on sample
geometry, and therefore does not reflect an {\em intrinsic} bulk
property of the vortex system.

As shown previously, the difference in the {\em dc} voltage
response of the two geometries is a result of the injection of a
metastable DP \cite{yosprl,yosna}. The DP, which is the {\em
equilibrium} phase above $H_{DT}^H$, can exist as a {\em
metastable} phase below $H_{DT}^H$
\cite{yosprl,maxim,yosna,hend96}. In the presence of a driving
current, vortices that penetrate through the rough edges locally
destroy the equilibrium OP, and form instead the {\em metastable}
DP near the edges. As the entire lattice flows deeper into the
sample, this inadvertently injected `wrong' phase anneals into the
OP over a characteristic relaxation length $L_r(H,I)$. The value
of $L_r$ is the key parameter that determines the voltage response
and the noise of the system. Since the DP has a significantly
larger critical current density $J_c^{dis}$ than $J_c^{ord}$ of
the OP, the contamination by the metastable DP enhances the
integrated $I_c$ of the strip: $I_{c}=d \int_0^WJ_c(x)dx =
WdJ_{c}^{ord}+ L_rd(J_{c}^{dis}- J_{c}^{ord})(1-exp(-W/L_r))$.
Here $d$ and $W$ are the thickness and width of the sample and we
have assumed, for simplicity, an exponential relaxation of $J_c(x)
= J_{c}^{ord}+ (J_{c}^{dis}- J_{c}^{ord})exp(-x/L_r)$. Since
usually experimentally $L_r < W$ \cite{yosna}, we can approximate
$I_{c}\simeq L_rd(J_{c}^{dis}- J_{c}^{ord})+ WdJ_{c}^{ord}$.
Importantly, $L_r(H,I)$ depends strongly on field as well as on
current. Close to the order-disorder transition fields $H_{DT}$
the free energies of the DP and OP are comparable; hence the
metastable DP has a long lifetime and $L_r$ is large \cite{yosna}.
This results in a large $I_c$ and consequently in almost vanishing
$V$ in the strip at fields slightly below $H_{DT}^H$ (Fig 1a). A
mirror-image-like behavior is observed in the vicinity of
$H_{DT}^L$ where $L_r$ increases again upon approaching the
reentrant transition from above. At intermediate fields, away from
the transitions, the metastable phase becomes less favorable
energetically and therefore $L_r$ shortens and $V$ is enhanced.

We now illustrate that the above edge contamination mechanism,
which accounts for both the $\it {dc}$ and the $\it {ac}$ response
of the system \cite{yosna}, also provides a compelling explanation
of the excess noise. The generation of the metastable DP results
from the non-uniform vortex penetration through the surface
barriers at the sample edge. The generation mechanism and the
subsequent annealing of the DP in the bulk constitute random
processes both in time and in space, resulting in complex
spatiotemporal fluctuations \cite{maslov} of the vortex-lattice
disorder and of the corresponding value of $J_c({\bf r},t)$. Two
mechanisms actually contribute to the spatiotemporal variations:
fluctuations of $J_c^{dis}$ at the sample edge, $\delta
J_c^{dis}$, and local fluctuations in the relaxation length
$\delta L_r$ within the bulk of the sample. Here we discuss for
simplicity only $\delta J_c^{dis}$ fluctuations, although
analogous conclusions are obtained on considering $\delta L_r$
fluctuations. In order to simplify the analysis we shall consider
a one-dimensional $J_c(x,t)$ problem, which effectively averages
out the sample properties along the current direction. Since
$J_c^{dis}$ is typically an order of magnitude larger than
$J_c^{ord}$ \cite{yosprl,hend96}, $\delta J_c^{dis}$ fluctuations
will result in large variations in $I_c$, causing large voltage
noise. The detailed microscopic properties of $\delta J_c^{dis}$
fluctuations are presently unknown. However, from the above
derivation of $I_c$, the fluctuation in the total critical current
caused by a given $\delta J_c^{dis}$ is $\delta I_c\simeq
L_r(H,I)\delta J_c^{dis}d$, which shows that $I_c$ fluctuations
strongly depend on the value of $L_r(H,I)$.

The low frequency voltage noise $\delta V$ can be evaluated as
follows. The {\it dc} voltage $V$, which appears at $I>I_c$, can
be expressed as $V=f(I-I_c)$, where $f$ is a general function
describing the {\it dc} $V-I$ characteristics. Therefore, $\delta
V= (df/dI_c)\delta I_c = - (df/dI)\delta I_c = - (dV/dI)\delta I_c
= -L_r(dV/dI) \delta J_c^{dis}d$. Consequently, as described
below, the commonly observed large excess noise in the lower part
of the PE, seen at about 3.5 kOe in Fig. 1b in the strip, is a
result of the large $L_r$ in the vicinity of $H_{DT}^H$. The novel
observation in Fig. 1b, however, is the existence of a second
noise peak at about 2 kOe where $L_r(H,I)$ becomes large again on
approaching $H_{DT}^L$. This low-field peak was not previously
observed since the noise studies were carried out on undoped
NbSe$_2$ \cite{marley95,merthew,rabin} which does not show a
reentrant disorder-driven transition. Our finding of the two peaks
is an important manifestation of the proposed mechanism: It
demonstrates that the excess noise is not a mere result of the
fact that $I_c$ increases with $H$ at the PE since at low fields
the same excess noise is found in the region where $I_c$ {\it
decreases} with $H$. Furthermore, the same value of $I_c$ is
attained at three values of $H$: above the reentrant $H_{DT}^L$
where $I_c$ decreases with $H$, below $H_{DT}^H$ where $I_c$
increases with $H$, and above $H_{DT}^H$ where $I_c$ decreases
again in the upper part of the PE. The excess noise occurs only in
the first two cases, where the metastable DP contaminates the
equilibrium OP. In the third case, above $H_{DT}^H$, the DP is the
thermodynamically stable phase and therefore no metastable phase
is generated at the edges.

In order to test the validity of the described concept we have
performed noise measurements in the Corbino geometry. Strikingly,
we find that the excess noise is entirely absent in Corbino in the
central-field region, as shown in Fig. 1b. This means that {\it
the motion of vortex lattice within the bulk of the sample does
not, by itself, create excess noise}. Any conventional bulk noise
mechanism should have resulted in a similar noise level in the
Corbino and strip geometries. One may even argue that some of the
noise mechanisms, such as a bulk plastic vortex flow, could cause
a larger noise level in the Corbino due to the enhanced vortex
shear by the $1/r$ radial current distribution, contrary to the
observations. The absence of the noise in the Corbino therefore
clearly indicates the dominant role of the edge contamination in
the noise process. The residual small and narrow peaks in the
Corbino noise in Fig. 1b can be ascribed to small deviations from
a perfect Corbino disk configuration. Since $L_r$ diverges at
$H_{DT}^H$ and $H_{DT}^L$, any small non-radial part of the
current or inhomogeneities may result in some injection of the
metastable DP, giving rise to noise. In the vicinity of the
mean-field $H_{DT}$ nonuniform disorder distribution may result in
some parts of the sample being in the equilibrium DP, whereas
others in the OP \cite{maxim}, similar to the solid-liquid
coexistence at melting \cite{sasha}. Consequently, when the
lattice is set in motion the DP drifts into regions of the OP,
where it becomes metastable, and may cause noise in this narrow
field region.

We now analyze the current dependence of the excess noise in the
strip. Figure 2 shows the low frequency spectral density of the
noise $S$(3 Hz) $\propto \delta V^2$ as a function of $I$, along
with the $V-I$ characteristic. The $V-I$ characteristic displays a
rapid upturn and approaches linear behavior with a constant
$dV/dI$ at elevated currents. $S$, in contrast, displays a large
peak and vanishes rapidly at higher currents. According to our
simplified analysis the voltage noise $\sqrt S \propto \delta V=
\delta J_c^{dis} L_r(H,I){dV \over dI}d$ is a product of three
separate terms. $\delta J_c^{dis}$ describes the statistical
process of the generation of the DP at the rough sample edges.
There is currently no theoretical description of this random
process which should generally be field and current dependent
\cite{note}. There is also no theoretical description of the
relaxation process, however, it is known experimentally that
$L_r(H,I)$ decreases rapidly with current \cite{yosna,hend96}.
When the lattice is displaced very slowly, the lifetime of the
metastable DP is long and $L_r$ is large. However, as the pinning
potential is tilted stronger by a larger driving force, the
annealing process becomes progressively faster \cite{hend96}
resulting in a rapid decrease of $L_r$ with vortex velocity.
Finally, even the third term $dV/dI$ is not well-defined
experimentally. Due to the metastable nature of the system $dV/dI$
has large fluctuations and strong dependence on measurement
frequency as well as on current ramp direction \cite{bhatta}.
These uncertainties in all three terms prevent quantitative
analysis of the noise amplitude at this stage. Nevertheless we can
understand the general form of the noise knowing the qualitative
behavior of $L_r(H,I)$ and $dV/dI$. The inset of Fig. 2 shows a
typical $dV/dI$ along with the noise intensity $\sqrt S$. $dV/dI$
shows a pronounced peak and reaches a constant value above about
30 mA. At low currents $L_r(H,I)$ is large and hence the noise
initially follows $dV/dI$. Above about 15 mA, however, the
increase of the noise is moderated due to the gradual decrease of
$L_r$. At still higher current the noise starts to drop rapidly
because of the rapid decrease of $L_r(H,I)$ with current. Above 28
mA, $dV/dI$ approaches a constant value of the flux-flow
resistance of the OP, indicating that $L_r$ is small and that most
of the sample volume is in the OP. Accordingly, noise decreases
rapidly in this region with decreasing $L_r(H,I)$. Since the
ordered part of the sample does not contribute to the noise, the
noise level vanishes as the width of the DP near the edge
diminishes.

>From the above considerations we can analyze the general noise
behavior of the strip, presented in Fig. 3. At the lowest current,
$I=$ 15 mA, vortex motion occurs only in the central field region
(Fig. 3a), since closer to $H_{DT}^H$ and $H_{DT}^L$ the
integrated $I_c$ of the strip is larger than 15 mA due to the
large $L_r$. As a result, the excess noise in Fig. 3b is present
only in the central part. At 18 mA, the field range of the
observable vortex motion and noise expands, and around 23 mA, two
noise peaks become apparent. In the central-field region the DP is
less stable, $L_r(H,I)$ drops with $I$, and hence the noise
decreases rapidly with the current. Closer to the transition
fields, however, the metastable DP is much more stable so that
$L_r$ remains large and $S$ still increases with current. At 36 mA
most of the sample is in the OP and the noise has accordingly
dropped by two orders of magnitude. The strong excess noise is
restricted now only to the narrow regions near $H_{DT}$ fields
where the metastable DP survives even at high vortex velocities.

In summary, the comparative study of excess noise generation in
Corbino and strip configurations shows that the flow of the vortex
lattice in the Corbino does not generate an excess voltage noise.
In contrast, very strong noise enhancement is found in the same
samples measured in a strip-like geometry. The excess noise is
found on the Bragg glass side of the disorder-driven transition
both along the high-field and the reentrant transition lines, and
results from random generation of a metastable disordered vortex
phase at the sample edges and its subsequent dynamic annealing in
the bulk.

We are grateful to D. E. Feldman and S. S. Banerjee for valuable
discussions. This work was supported by the Israel Science
Foundation - Center of Excellence Program and by the US-Israel
Binational Science Foundation (BSF). EZ acknowledges support by
the German-Israeli Foundation G.I.F. and by the Fundacion
Antorchas - WIS program.


\ FIGURE CAPTIONS

Fig. 1. (a) The voltage response $V$ vs. magnetic field at 4.4 K measured in
the Corbino ($\bigcirc $) and strip ($\bullet $) geometries. Inset: same
data over a wider field range extending above $H_{c2}$. (b) The
corresponding Corbino ($\bigcirc $) and strip ($\bullet $) noise power at 9
Hz showing the absence of the excess noise in the Corbino at intermediate
fields. Inset: the electrode configuration allowing measurements in both the
Corbino and strip configurations.

Fig. 2. The $V-I$ characteristic ($\bullet $) and the noise power density at 3 Hz
($\bigcirc $) in the strip configuration. Inset: representative noise power
spectra at various fields in the vicinity of $H_{DT}^L$.

Fig. 3. (a) Voltage response vs. field at 4.2 K in the strip configuration
for $I=$ 36 mA ($\square $), 23 mA ($\bigcirc $), 18 mA ($\blacksquare $)
and 15mA ($\bullet $). (b) The corresponding noise power density at 3 Hz.

\end{multicols}

\begin{references}
\bibitem{note} $\delta J_c^{dis}$ and the resulting $\delta V$ also
have nontrivial $1/f$-like spectral form which will be addressed
in detail elsewhere.
\end{references}
\end{document}